\pdfoutput=1

\documentclass[journal]{IEEEtran}
\ifCLASSOPTIONcompsoc
  \usepackage[nocompress]{cite}
\else
  \usepackage{cite}
\fi
\ifCLASSINFOpdf
  \usepackage[pdftex]{graphicx}
\else
\fi
\usepackage{amsmath}
\usepackage{booktabs}
\usepackage{multirow}
\newcommand{\etal}{{\em et al\,.}}       
\newcommand{\eg}{{e.g.}}           
\newcommand{\ie}{{i.e.}}           

\begin{document}
%
\title{Synergistic Learning of Lung Lobe Segmentation and Hierarchical Multi-Instance Classification for Automated Severity Assessment of COVID-19 in CT Images}
%
%
%
%

\author{Kelei He,~Wei Zhao,~Xingzhi Xie,~Wen Ji,~Mingxia Liu,~Zhenyu Tang,~Feng Shi,~Yang Gao\\Jun Liu,~Junfeng Zhang,~Dinggang Shen,~\IEEEmembership{Fellow,~IEEE}
\thanks{This work is supported in part by Key Emergency Project of Pneumonia Epidemic of novel coronavirus infection under grant 2020sk3006, Emergency Project of Prevention and Control for COVID-19 of Central South University under grant 160260005, Foundation of Changsha Scientific and Technical Bureau under grant kq2001001, and National Key Research and Development Program of China under grant 2018YFC0116400. Corresponding authors: Jun Liu (junliu123@csu.edu.cn); Junfeng Zhang (jfzhang@nju.edu.cn); Dinggang Shen (Dinggang.Shen@gmail.com).}
\thanks{K. He and W. Zhao contributed equally to this work.}
\thanks{K. He and J. Zhang are with Medical School of Nanjing University, Nanjing, China. K. He, Y. Gao and J. Zhang are also with the National Institute of Healthcare Data Science at Nanjing University, China (e-mail: hkl@nju.edu.cn).}
\thanks{W. Zhao, X. Xie and J. Liu are with the Department of Radiology, the Second Xiangya Hospital, Central South University, Changsha, Hunan, China. J. Liu is also with the Department of Radiology Quality Control Center, Changsha, Hunan, China (e-mail: wei.zhao@csu.edu.cn; xingzhixie123@csu.edu.cn).}
\thanks{W. Ji and Y. Gao are with State Key Laboratory for Novel Software Technology, Nanjing University, Nanjing, China (e-mail: jiwen@smail.nju.edu.cn; gaoy@nju.edu.cn).}
\thanks{M. Liu is with Biomedical Research Imaging Center and the Department of Radiology, University of North Carolina, Chapel Hill, NC, U.S. (e-mail: mxliu@med.unc.edu).}
\thanks{Z. Tang is with Beijing Advanced Innovation Center for Big Data and Brain Computing, Beihang University, Beijing, China (e-mail: tangzhenyu@buaa.edu.cn).}
\thanks{F. Shi and D. Shen are with the Department of Research and Development, Shanghai United Imaging Intelligence Co., Ltd., Shanghai 200232, China (e-mail: feng.shi@united-imaging.com).}}

%
%

\markboth{Journal of \LaTeX\ Class Files,~Vol.~14, No.~8, August~2015}%
{K. He \MakeLowercase{\textit{et al.}}: Synergistic Learning for Automated Severity Assessment of COVID-19 in CT Images}
%



\IEEEtitleabstractindextext{%
\begin{abstract}
Understanding chest CT imaging of the coronavirus disease 2019 (COVID-19) will help detect infections early and assess the disease progression.
Especially, automated severity assessment of COVID-19 in CT images plays an essential role in identifying cases that are in great need of intensive clinical care. However, it is often challenging to accurately assess the severity of this disease in CT images, due to variable infection regions in the lungs, similar imaging biomarkers, and large inter-case variations. To this end, we propose a synergistic learning framework for automated severity assessment of COVID-19 in 3D CT images, by jointly performing lung lobe segmentation and multi-instance classification. Considering that only a few infection regions in a CT image are related to the severity assessment, we first represent each input image by a bag that contains a set of 2D image patches (with each cropped from a specific slice). A multi-task multi-instance deep network (called M$^2$UNet) is then developed to assess the severity of COVID-19 patients and also segment the lung lobe simultaneously. Our M$^2$UNet consists of a patch-level encoder, a segmentation sub-network for lung lobe segmentation, and a classification sub-network for severity assessment (with a unique hierarchical multi-instance learning strategy). Here, the context information provided by segmentation can be implicitly employed to improve the performance of severity assessment. Extensive experiments were performed on a real COVID-19 CT image dataset consisting of 666 chest CT images, with results suggesting the effectiveness of our proposed method compared to several state-of-the-art methods.
\end{abstract}

}

\maketitle

\IEEEdisplaynontitleabstractindextext

%
\IEEEpeerreviewmaketitle


%
%
%
%


\section{Introduction}
\label{sec:introduction}
\IEEEPARstart{T}{HE} coronavirus disease 2019 (COVID-19) is spreading fast worldwide since the end of $2019$. Until April 5, about $1.03$ million patients are confirmed with this infectious disease, reported by~\cite{pubcode}. This raises a Public Health Emergency of International Concern (PHEIC) of WHO. In the field of medical image analysis, many imaging-based artificial intelligence methods have been developed to help fight against this disease, including automated diagnosis~\cite{chen2020deep,song2020deep,tang2020severity}, ~segmentation~\cite{shan+2020lung,jin2020ai,qi2020machine}, and follow-up and prognosis~\cite{ReviewCOV}.

\begin{figure}[!t]
\setlength{\abovecaptionskip}{0pt}
  \centering
  \includegraphics[width=0.9\linewidth]{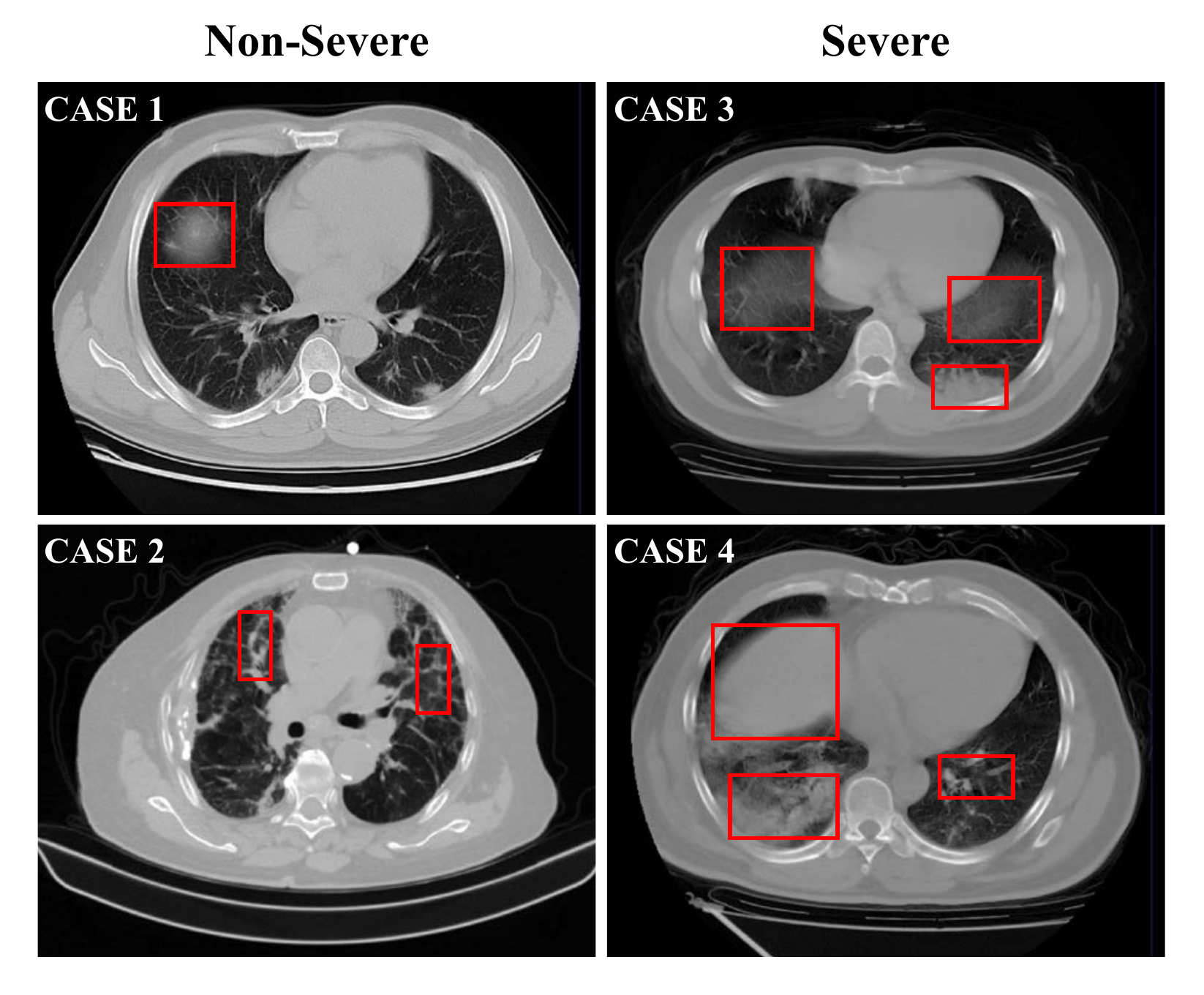}
    \caption{\label{fig:spotlight}Typical cases of two non-severe (left) and two severe (right) patients with COVID-19, where infections often occur in small regions of the lungs in CT images. The similar imaging biomarkers (\eg, ground glass opacities, mosaic sign, air bronchogram and interlobular septal thickening) of both cases (denoted by red boxes) make the non-severe and severe images difficult to distinguish.}
\end{figure}

\begin{figure*}[!t]
\setlength{\abovecaptionskip}{0pt}
  \centering
  \includegraphics[width=0.95\linewidth]{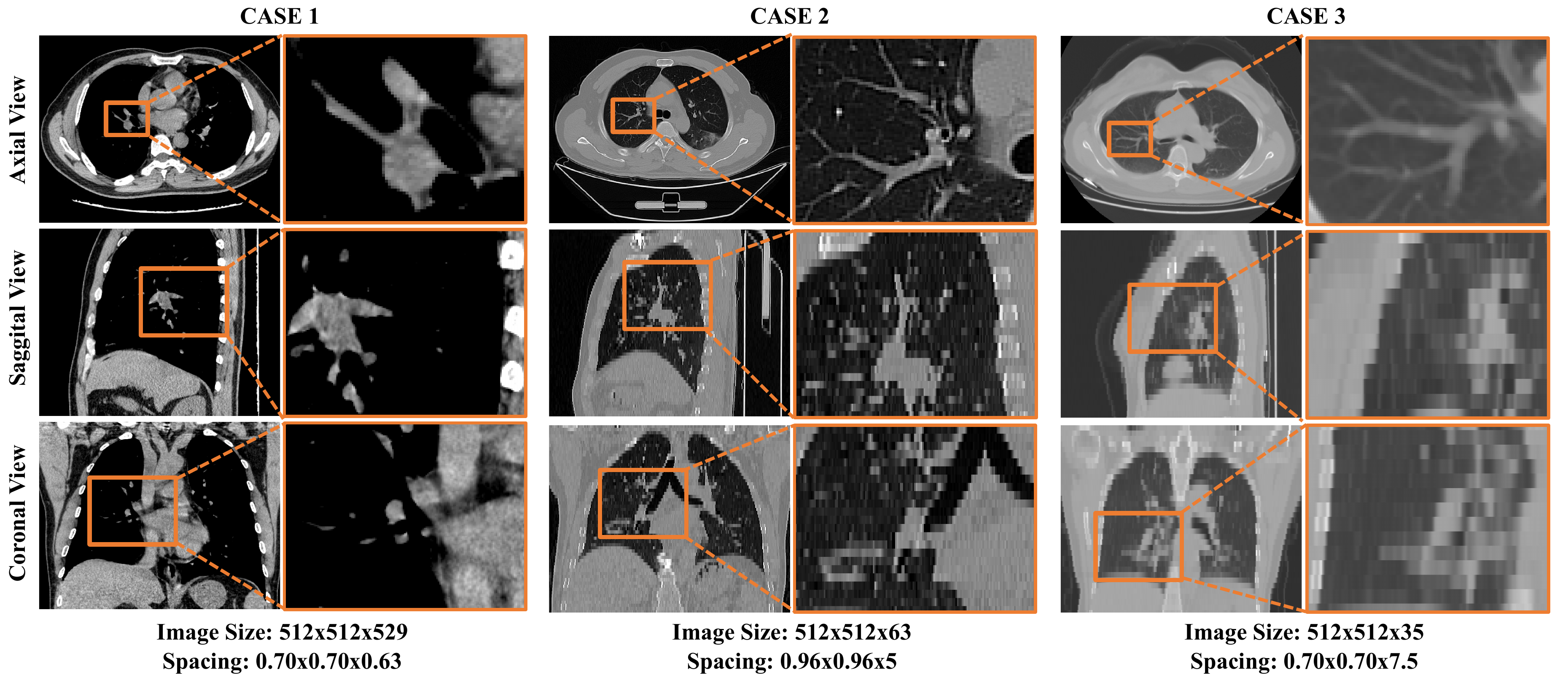}
    \caption{\label{fig:image_compare} Visualization of three typical cases in the COVID-19 CT image dataset from three different views. As shown in this figure, large inter-case variations (\eg, image size and spatial resolution) exist in CT images of COVID-19 patients.}
\end{figure*}

Previous imaging-based studies mainly focus on identifying COVID-19 patients from non-COVID-19 subjects. As the golden standard for COVID-19 is the Reverse Transcription-Polymerase Chain Reaction (RT-PCR) test, the effectiveness of those imaging-base applications is limited. 

Moreover, approximately $80\%$ of patients with COVID-19 have only mild to moderate symptoms~\cite{verity2020estimates}, while the remaining patients have severe symptoms. 
Based on previous studies \cite{xie2020chest,shan+2020lung}, the imaging-based characters of COVID-19 patients are distinct to related diseases, \eg, viral pneumonia. Therefore, the severity assessment of the disease is of high clinical value, which helps effectively allocate medical resources such as ventilator. 
Among various radiological examinations, chest CT imaging plays an essential role in fighting this infectious disease by helping early identify lung infections and assess the severity of the disease. Previous studies show that computed tomography (CT) has the ability to provide valuable information in the screening and diagnosis \cite{xie2020chest}. In this work, CT help the clinicians to evaluate the condition of the patients in advance, by which necessary measures or treatments could better proceed, especially for severe patients in time.

However, automatically assess the severity of COVID-19 in CT images is a very challenging task. 
First, infections caused by COVID-19 often occur in small regions of the lungs and are difficult to identify in CT images, as shown in Fig.~\ref{fig:spotlight}.
Second, imaging biomarkers of COVID-19 patients caused by an infection are similar in some severe and non-severe cases are similar, including ground-glass opacities (GGO), mosaic sign, air bronchogram, and interlobular septal thickening (Fig.~\ref{fig:spotlight}). 
In addition, there are large inter-case variations in CT images of COVID-19 patients (Fig. \ref{fig:image_compare}), because these images are usually acquired by multiple imaging centers with different scanners and different scanning parameters.

Several recent methods have been proposed for the diagnosis of COVID-19~\cite{chen2020deep,song2020deep,tang2020severity}, with only some specifically designed for severity assessment of the disease. In several studies ~\cite{shan+2020lung,jin2020ai,qi2020machine}, segmentation of lung or lung lobe is used as a prerequisite procedure for diagnosis purposes.

However, most of these methods treat the lung lobe segmentation and disease diagnosis as two separate tasks, ignoring their underlying association. Note that the segmentation of lung lobe can provide rich information regarding spatial locations and tissue types in CT images. Therefore, it is intuitively reasonable to jointly perform lung lobe segmentation and severity assessment/prediction, where the context information provided by segmentation results can be used to improve the prediction performance. The joint learning scheme is obviously faster than the two-stage framework, since detecting and cropping the lung field are not needed. Besides, as demonstrated by related works of class activation maps (CAMs)~\cite{zhou2016learning}, the classification task raises high signal responses in lung lobe area, where the infection patterns of lung lobe in disease progression could also provide useful guidance for the segmentation of lung lobes.

Moreover, most of the previous works are based on 2D image slices~\cite{chen2020deep,song2020deep,tang2020severity}. 
However, the annotation of 2D CT slices is a heavy workload for radiologists.
It is interesting to directly employ 3D CT images for automated severity assessment of COVID-19, which is desired for real-world clinical applications.

To this end, in this paper, we propose a synergistic learning framework for automated severity assessment of COVID-19 in 3D CT images, by jointly performing severity assessment of COVID-19 patients and lung lobe segmentation.
The proposed method extends the previous conference work \cite{he2019midcn} to deal with 3D images, and makes substantial methodological improvements for the task of COVID-19 severity assessment.
Specifically, considering that only a few slices in CT images are related to severity assessment, each input CT image is represented by a bag of 2D image patches, each of which is randomly cropped from a specific slice. Furthermore, each slice is represented by a bag of infection regions represented by intermediate embedding features. With each bag as input, a multi-task multi-instance deep neural network (called M$^2$UNet) is developed, including 1) a shared patch-level encoder, 2) a segmentation sub-network for lung lobe segmentation, and 3) a classification sub-network for severity assessment of COVID-19 patients (\ie, severe or non-severe) using a hierarchical multi-instance learning strategy. Here, the segmentation results are used to provide context information of input CT images to boost the performance of severity assessment. Extensive experiments have been performed on a real-world COVID-19 dataset with $666$ chest CT images, with the results demonstrating the effectiveness of the proposed method compared to several state-of-the-art methods.

The contributions of this work are three-fold:
\begin{itemize}
    \item A multi-task multi-instance learning framework is proposed to jointly assess the severity of the COVID-19 patients and segment lung lobes in chest CT images, where the segmentation task provides context information to aid the task of severity assessment in chest CT image.
    \item A unique hierarchical multi-instance learning strategy is developed to predict the severity of patients in a weakly supervised manner.
    \item We evaluate the proposed method on a real clinical dataset with 666 3D CT images of COVID-19 patients, achieving promising results in severity assessment compared to several state-of-the-art methods.
\end{itemize}

The rest of the paper is organized as follows. In Section~2, we introduce the related works for the segmentation and diagnosis of CT images of COVID-19 patients, as well as related studies on deep multi-instance learning. Then, we introduce the proposed method in Section~3. In Section~4, we present the materials, experimental setup, and experimental results. 
Finally, we conclude this paper and present several future research directions in Section~5.

\section{Related Work}
In this section, we briefly review the most relevant studies from the following three aspects: 1) lung segmentation of CT images with COVID-19, 2) automated diagnosis of COVID-19 patients, and 3) deep multi-instance learning.

\subsection{Lung Segmentation of CT Images with COVID-19}
Segmentation of lung or lung lobe has been used as a common pre-requisite procedure for automatic diagnosis of COVID-19 based on chest CT images. Several deep learning methods have been proposed for the segmentation of lung in CT images with COVID-19. For instance, U-Net~\etal \cite{ronneberger2015u} has been widely used for segmentation of both lung regions and lung lesions in COVID-19 applications~\cite{zheng2020deep,cao2020longitudinal,huang2020serial,qi2020machine}. Qi~\etal~\cite{qi2020machine} use U-Net to delineate the lesions in the lung and extract radiometric features of COVID-19 patients with the initial seeds given by a radiologist for predicting hospital stay. Also, several variants of U-Net have been applied to the diagnosis or severity assessment of COVID-19. Jin~\etal~\cite{jin2020ai} design a two-stage pipeline to screen COVID-19 in CT images, and they utilize U-Net++~\cite{zhou2018unet++} to detect the whole lung region and to separate lesions from lung regions. Besides, V-Net~\cite{milletari2016v} is also used in various segmentation applications. Shan~\etal~\cite{shan+2020lung} integrates human-in-the-loop strategy into the training process of VB-Net (a variant of V-Net). The human-aided strategy is an intuitive way to address the issue of lacking manual labels during segmentation in CT images.

\subsection{Automated Diagnosis of COVID-19}
Both X-rays \cite{wong2020frequency} and CT images \cite{xie2020chest} can provide effective information for the computer-assisted diagnosis of COVID-19. Compared with X-rays, chest CT imaging contains hundreds of slices, which is clearer and more precise but has to take more time for specialists to diagnose. Therefore, there is a great demand to use CT images for automated diagnosis of COVID-19. In general, the existing methods for COVID-19 diagnosis based on CT images can be roughly divided into two categories: 1) classification; 2) severity assessment. In the former category, many studies have been conducted to determine whether patients are infected with COVID-19 disease. For example, Chen~\etal~\cite{chen2020deep} exploits a UNet++ based segmentation model to segment COVID-19 related lesions in chest CT images of 51 COVID-19 patients and 55 patients with other diseases, and finally determine the label (COVID-19 or non-COVID-19) of each image based on the segmented lesions. Ying~\etal~\cite{song2020deep} propose a CT diagnosis system, namely DeepPneumonia, which is based on the ResNet50 model to identify patients with COVID-19 from bacteria pneumonia patients and healthy people. In the second category, Tang~\etal \cite{tang2020severity} proposed to first adopt VB-Net to separate the lung into anatomical sub-regions, and then use these sub-regions to compute quantitative features for training a random forest (RF) model for COVID-19 severity assessment (with labels of being non-severe or severe).

\begin{figure*}[!t]
  \centering
  \includegraphics[width=\linewidth]{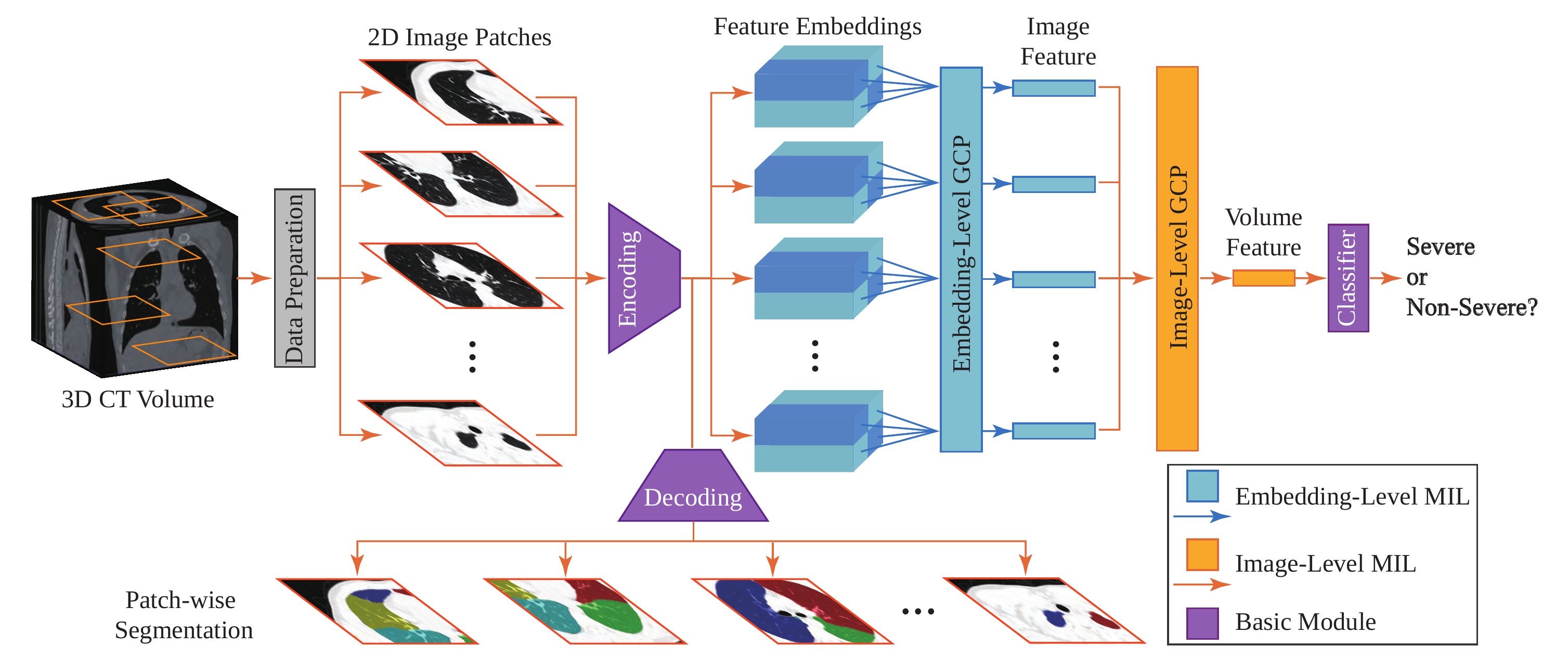}
    \caption{\label{fig:framework} Illustration of the proposed framework for joint lung lobe segmentation and severity assessment of COVID-19 in 3D CT images. Each raw 3D CT image is first pre-processed, and multiple 2D image patches (with each patch from a specific slice) are then extracted to construct an instance bag for representing each input CT scan. This bag is then fed into the proposed multi-task multi-instance UNet (M$^2$UNet) for joint lung lobe segmentation and severity assessment of COVID-19, consisting of a shared patch-level encoder, a segmentation sub-network, and a classification sub-network for severity assessment. Here, the segmentation task can provide location and tissue guidance for the task of severity assessment that employs a hierarchical multi-instance learning strategy.}
\end{figure*}

\subsection{Deep Multi-instance Learning}
The scenario of multi-instance learning (MIL)~\cite{dietterich1997solving,xu2015show,liu2018joint} or learning from weakly annotated data~\cite{oquab2014weakly} arises when only a general statement of the category is given, but multiple instances can be observed. MIL aims to learn a model that can predict the label of a bag accurately, and many recent studies have focused on implementing MIL via deep neural networks. 
For instance, Oquab~\etal~\cite{oquab2014weakly} train a deep model with multiple image patches of multiple scales as input, and aggregate the prediction results of multiple inputs by using a max-pooling operation. Besides, many studies~\cite{feng2017deep,sun2016multiple,wu2015deep} propose to formulate image classification as a MIL problem so as to address the weakly supervised problem. Moreover, MIL is particularly suitable for problems with only a limited number (\eg, tens or hundreds) of training samples in various medical image-based applications, such as computer-assisted disease diagnosis~\cite{tong2014multiple,xu2014weakly,yan2016multi,liu2018landmark}. For instance, Yan~\etal \cite{yan2016multi} propose a two-stage deep MIL method to find discriminative local anatomies, where the first-stage convolutional neural network (CNN) is learned in a MIL fashion to locate discriminative image patches and the second-stage CNN is boosted using those selected patches. More recently, a landmark-based deep MIL framework~\cite{liu2018landmark} is developed to learn both local and global representations of MRI for automated brain disease diagnosis, leading to a new direction for handling limited training samples in the domain of medical image analysis. Since there are only a limited number of cases at hand, it is desirable to employ the multi-instance learning strategy for severity assessment of COVID-19 patients in chest CT images.

\section{Proposed Method}

\subsection{Framework}
The framework of the proposed method is illustrated in Fig.~\ref{fig:framework}, where the input is the raw 3D CT image and the output is the lung segmentation and severity assessment of COVID-19 patients (\ie, severe or non-severe). Specifically, each 3D CT image is processed via several image pre-processing steps. Then, a set of 2D image patches is randomly cropped from the processed image to construct an instance bag, and each bag represents a specific input CT image. This bag is regarded as the input of the proposed multi-task multi-instance U-Net (M$^2$UNet). The M$^2$UNet is designed to learn two tasks jointly, \ie, severity assessment of a COVID-19 patient and segmentation of the lung lobe. 

As shown in Fig.~\ref{fig:framework}, in M$^2$UNet, an encoding module is first used for patch-level feature extraction of 2D patches in each input bag, followed by two sub-networks for joint severity assessment and lung lobe segmentation. Specifically, in the classification sub-network, these extracted patch-level features are fed into a feature embedding module and an image-level feature learning module to capture the local-to-global volume representation of the input CT image. With the learned volume features, a classification layer is finally used to assess the severity of each COVID-19 patient (\ie, severe or non-severe). 
In the segmentation sub-network, those patch-level features are fed into a decoding module to perform lung lobe segmentation for each patch in the input bag. 
Since these two sub-networks are trained jointly with a shared patch-level encoder, the context information provided by the segmentation results can be implicitly employed to improve the performance of severity assessment.

\subsection{Data Preparation}
\label{S:DataPre}
To eliminate the effect of the background noise in each raw 3D CT image, we crop each scan to only keep the region containing the human body, by using a threshold-based processing method. Specifically, we first binarize the image using the threshold of zero, through which the human tissues and the gas regions will be separated. Then, the human body region is cropped according to the binary mask. Each body region image has a size of at least $256\times256$ for the axial plane in this work.

While image resampling is commonly used in many deep learning methods for segmentation and classification \cite{he2018pelvic,lian2018hierarchical}, we do not resample the raw CT images in order to preserve their original data distributions. Since our method is clinical-oriented with inconsistent imaging qualities, CT images used in this study are not as clean as those in benchmark datasets~\cite{bakas2018identifying}. For example, the physical spacing of our data has large variation, \eg, from $0.6\,mm$ to $10\,mm$ between slices, because of the use of different CT scanners and scanning parameters. Using a common interpolation method (\eg, tri-linear interpolation) to resample a CT image into $1\,mm$, one will introduce heavy artifacts to the image. Besides, only a few infection regions in each CT image are related to severity assessment. To this end, we employ the weakly supervised multi-instance learning (MIL) strategy for handing these inconsistent CT images. Specifically, for each pre-processed CT image, we randomly crop a set of 2D patches sampled from 2D slices (with each patch from a specific slice) in each image to construct an instance bag, and each bag is used to represent a specific CT image and treated as the input of the subsequent M$^2$UNet. In this way, the inter-slice/patch relationships can be implicitly captured by our M$^2$UNet. 
In addition, this MIL strategy represents each 3D image through a set of 2D image patches rather than sequential slices. This can partially alleviate the problem of data inconsistency, so our method has high practical value in real-world applications.

\subsection{Network Architecture}
As shown in Fig.~\ref{fig:framework}, using each bag (consisting of a set of 2D image patches) as the input data, the proposed M$^2$UNet first employs an encoding module for patch-level feature extraction. Based on these features, the classification and segmentation sub-networks are then used to jointly perform two tasks, respectively, \ie, 1) severity assessment of the patients, and 2) segmentation of lung lobes in each patch. Specifically, the classification sub-network uses a unique hierarchical MIL strategy to extract the local-to-global representation of each input image, with an embedding-level MIL module, an image-level MIL module, and a classification module. The segmentation sub-network contains a decoding module to segment lung lobes of 2D image patches in each bag.

The detailed network architecture is listed in Table \ref{Table:NetArc}. The combination of the encoder and decoder is U-Net like, with four down-sampling blocks in the encoder and four up-sampling blocks in the decoder. The outputs of the same level blocks in the encoder and decoder are concatenated and fed into the next block of the decoder. Limited by computational resources, all the convolutional layers in the encoder and decoder have the same number (\ie, $64$) of kernels, except the last block in the encoder. The last block of encoder outputs $512$ dimensional features to help build a more robust classification for severity assessment. The decoder outputs the corresponding segmentation mask of five types of lung lobes for each image patch.

\begin{table*}[htbp]
\renewcommand{\arraystretch}{1.1}
\centering
\caption{\label{Table:NetArc}Network architecture of the proposed M$^2$UNet. The network has three main components: 1) a encoding module containing five encoding blocks; 2) a classification sub-network containing the embedding-level MIL and image-level MIL, and a classifier; and 3) a segmentation sub-network consisting of a decoding module with five decoding blocks. MIL: multi-instance learning; Num.: Number of layers, K: kernel size; PAD: padding size; STR: stride; \#: Number of learnable parameters; cov: convolution; GCP: global contrast pooling; concat: concatenation.}
\begin{tabular}{|c|c|c|c|c|c|}
\toprule[1pt]
\textbf{Block Name} &\textbf{Num.} & \textbf{Layers} & \textbf{Parameter Setting} & \textbf{Input} & \textbf{\#}\\
\toprule[1pt]
Encoding block 1 & 2 & \{conv, batchnorm, ReLU\} & K: \{$3\times3\times64$\}, PAD:1, STR:1 & 2D image patches & $37K$ \\
\hline
Pool 1 & 1 & max-pooling & K: \{$2\times2$\}, STR:2 & Encoding block 1 & $-$\\
\hline
Encoding block 2 & 2 & \{conv, batchnorm, ReLU\} & K: \{$3\times3\times64$\}, PAD:1, STR:1 & Pool 1 & $72K$\\
\hline
Pool 2 & 1 & max-pooling & K: \{$2\times2$\}, STR:2 & Encoding block 2 & $-$\\
\hline
Encoding block 3 & 2 & \{conv, batchnorm, ReLU\} & K: \{$3\times3\times64$\}, PAD:1, STR:1 & Pool 2 & $72K$\\
\hline
Pool 3 & 1 & max-pooling & K: \{$2\times2$\}, STR:2 & Encoding block 3 & $-$\\
\hline
Encoding block 4 & 2 & \{conv, batchnorm, ReLU\} & K: \{$3\times3\times64$\}, PAD:1, STR:1 & Pool 3 & $72K$\\
\hline
Pool 4 & 1 & max-pooling & K: \{$2\times2$\}, STR:2 & Encoding block 4 & $-$\\
\hline
\multirow{2}{*}{Encoding block 5}& 1 & \{conv, batchnorm, ReLU\} & K: \{$3\times3\times64$\}, PAD:1, STR:1 & \multirow{2}{*}{Pool 4} & \multirow{2}{*}{$2595K$}\\
& 1 & \{conv, batchnorm, ReLU\} & K: \{$3\times3\times512$\}, PAD:1, STR:1 & & \\
\hline
\multirow{2}{*}{Embedding-Level MIL}& 1 & GCP & Num. Concepts: 256 & \multirow{2}{*}{Encoding block 5} & \multirow{2}{*}{$193K$}\\
& 1 & conv & K: \{$1\times1\times256$\}, PAD:0, STR:1 & & \\
\hline
\multirow{2}{*}{Image-Level MIL}& 1 & GCP & Num. Concepts: 128 & \multirow{2}{*}{Embedding-Level MIL} & \multirow{2}{*}{$48K$}\\
& 1 & conv & K: \{$1\times1\times128$\}, PAD:0, STR:1 & & \\
\hline
Clssifier & 1 & conv & K: \{$1\times1\times128$\}, PAD:0, STR:1 & Image-Level MIL & $0.3K$ \\
\hline
\multirow{2}{*}{Decoding block 5}& 1 & \{up-sample, conv, batchnorm, ReLU, concat\} &  K: \{$3\times3\times512$\}, PAD:1, STR:1 & Encoding block 5 & \multirow{2}{*}{$397K$}\\
& 2 & \{conv, batchnorm, ReLU\} & K: \{$3\times3\times128$\}, PAD:1, STR:1 & Encoding block 4& \\
\hline
\multirow{2}{*}{Decoding block 4}& 1 & \{up-sample, conv, batchnorm, ReLU, concat\} & K: \{$3\times3\times64$\}, PAD:1, STR:1 & Decoding block 5 & \multirow{2}{*}{$145K$}\\
& 2 & \{conv, batchnorm, ReLU\} & K: \{$3\times3\times128$\}, PAD:1, STR:1 & Encoding block 3 & \\
\hline
\multirow{2}{*}{Decoding block 3}& 1 & \{up-sample, conv, batchnorm, ReLU, concat\} & K: \{$3\times3\times64$\}, PAD:1, STR:1 & Decoding block 4 & \multirow{2}{*}{$145K$}\\
& 2 & \{conv, batchnorm, ReLU\} & K: \{$3\times3\times128$\}, PAD:1, STR:1 & Encoding block 2 & \\
\hline
\multirow{2}{*}{Decoding block 2}& 1 & \{up-sample, conv, batchnorm, ReLU, concat\} & K: \{$3\times3\times64$\}, PAD:1, STR:1 & Decoding block 3 & \multirow{2}{*}{$145K$}\\
& 2 & \{conv, batchnorm, ReLU\} & K: \{$3\times3\times128$\}, PAD:1, STR:1 & Encoding block 1 & \\
\hline
Decoding block 1 & 1 & conv & K: \{$1\times1\times64$\}, PAD:0, STR:1 & Decoding block 2 & $0.5K$\\
\hline
\end{tabular}
\end{table*}

\begin{figure}[!th]
  \centering
  \includegraphics[width=0.9\linewidth]{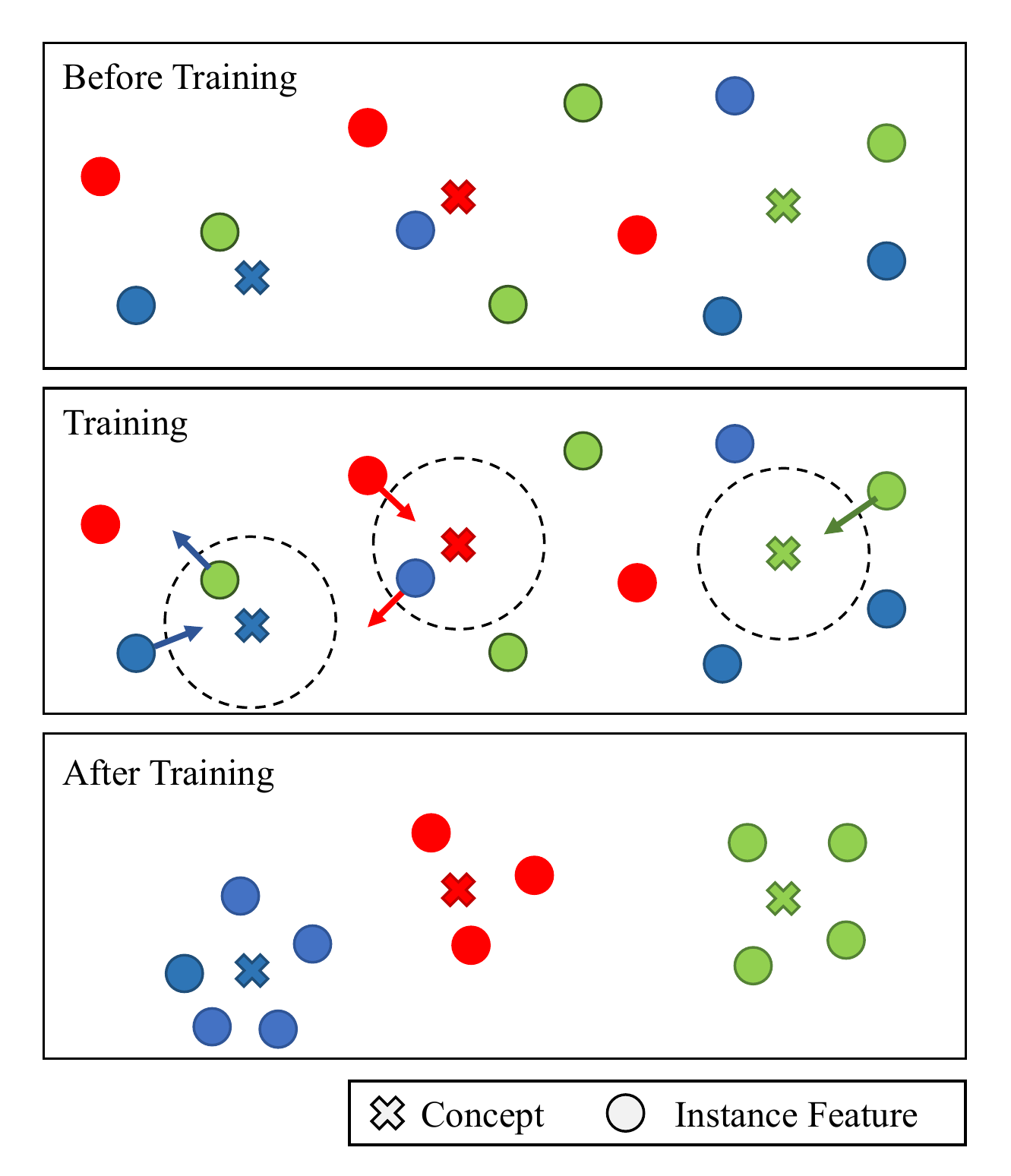}
    \caption{\label{fig:gcp} A brief illustration of the learning principle for the proposed global contrast pooling (GCP) layer. Here, the concepts denote to-be-learned features that are discriminative for severity assessment. The GCP layer is designed to pull the relevant instance features and concepts closer, and push the irrelevant instance features and concepts away from each other.}
\end{figure}

\subsection{Hierarchical Multi-Instance Learning}
While infection regions of the lung, related to COVID-19 (\eg, nodule and GGO) are usually located in regions of the CT image, the category of each CT image is labeled at the entire image level, rather than the region-level. That is, many regions are actually unrelated to the classification task for severity assessment. 

Multi-instance learning (MIL) provides a useful tool to solve such a weakly supervised problem. 
Conventional MIL represents a 2D image as a bag, and each bag consists of multiple regions of the input image (\ie, instances). Their overall prediction is made at the bag-level by roughly two kinds of methods, \ie, the embedding-level methods and the instance-level methods. The former learns the relationship among instances by projecting them into a new embedding space. The latter directly generates the bag-level predictions by performing voting on the instance predictions. However, both methods are inapplicable for the classification of 3D images, as 3D images contain multiple 2D image slices, and the class labels are also related to some local regions of the slices.

In this work, based on the previous study on pathological images \cite{he2019midcn}, we propose a hierarchical MIL strategy in the classification sub-network of our M$^2$UNet to perform severity assessment of COVID-19 in 3D CT images, as shown in Fig.~\ref{fig:framework}.
As mentioned in Section~\ref{S:DataPre}, we represent each input 3D volumetric image as a bag consisting of a set of 2D image patches, and these patches are regarded as the instances in the MIL problem settings. 
Formally, we first construct a bag with $n$ 2D patches cropped from the regional slices to represent each input CT image. Denote the $i$-th and the $j$-th 3D CT image as ${\mathcal{X}_i}$ and ${\mathcal{X}_j}$, respectively, where $\mathcal{X}_i=\{\phi_{i1}^{ins}, \phi_{i2}^{ins}, \cdots, \phi_{in_i}^{ins}\}$ and $\mathcal{X}_j=\{\phi_{j1}^{ins}, \phi_{j2}^{ins}, \cdots, \phi_{jn_j}^{ins}\}$. Here, $\phi_{kl}^{ins}\in \mathrm{R}^d$ $(k = 1,2,\cdots,n_k)$ indicates the $l$-th instance of the $k$-th image.
Then, these 2D patches (size: $height \times width$) are fed into the encoding module for patch/instance-level feature extraction. These instance-level features are further fed into an embedding-level MIL module, which will be introduced later. After obtaining the instance-level features, the bag/image-level feature $\Phi_i$ are then generated by our proposed global contrast pooling (GCP) layer in the image-level MIL module.

As illustrated in Fig.~\ref{fig:gcp}, the proposed GCP layer aims to make the instance features closer to the relevant concepts, and also push those irrelevant instance features and concepts away from each other. In this work, the term `concept' denotes the to-be-learned feature of GCP layer that is discriminative for severity assessment. Theoretically, the concept is a normalized weight to map features in instance feature space to an ordered embedding space. Specifically, in the GCP layer, we assume the bag-level feature $\Phi_i$ is represented by the relationship between instance features and $p$ concepts. Here, these concepts are learned to reveal the data structure in a global perspective. The bag-level feature is then denoted as a $p$ dimensional feature vector, with each dimension denoting the maximum similarity between one concept and all instance features. We use the cosine function to measure such relationships. Thus, the bag feature and the similarity can be written as
\begin{equation}
\mathbf{\Phi}_{i}=[s_{i1},s_{i2},\cdots,s_{im},\cdots,s_{ip}],
\label{eq_Phi}
\end{equation}

\begin{equation}
s_{im}=\max_{k=1}^{n_i}\mathbf{w}_{m}^{\top}\mathbf{\phi}_{ik}+\mathbf{R}(\mathbf{w}_m),
\label{eq_sim}
\end{equation}

where $s_{im}$ ($m=1,\cdots,p$) is the maximum similarity between the instance features of the $i$-th bag and the $m$-th image-level concept $w_m$. $R(\cdot)$ denotes the commonly used regularization term used in deep networks. With Eqs.~\eqref{eq_Phi}-\eqref{eq_sim}, one can observe that the proposed GCP layer can automatically learn the concepts that are related to those discriminative instances, thus reducing the influence of those irrelevant instances. Note such a GCP layer can be also used in other weakly supervised problems, where only a small portion of regions in an image are related to the task at hand (such as MRI-based brain disorder diagnosis~\cite{liu2018landmark}).

We further use an embedding-level MIL module (with a GCP layer) to learn embedding-level representations, by regarding each image patch as a bag and the intermediate patch-level features produced by the encoder as instances. In this way, the relationships among small regions in each patch can be modeled by our method. Based on the embedding-level features, an image-level MIL module (with a GCP layer) is further used to generate the volume features. 
Based on the volume features, we use a fully-connected layer followed by a cross-entropy loss to predict the severity score (\ie, severe or non-severe) of each input CT image. The final loss function in the proposed hierarchical MIL network for severity assessment can be formulated as

\begin{equation}
\begin{array}{rrclcl}
\mathcal{L}_{MIL} = - log(fc(\Phi_i), y),
\end{array}
\label{eq:mil}
\end{equation}

where $fc(\cdot)$ denotes the mapping function of the fully-connected layer, and $y$ denotes the severity type confirmed by clinicians.

\subsection{Multi-task Learning for Joint Lung Segmentation and Severity Assessment}

The segmentation task is supervised by the aggregation of cross-entropy loss and Dice loss as follows
\begin{equation}
\begin{array}{rrclcl}
\mathcal{L}_{seg} = - \frac{1}{N} \sum \limits^{N}_{n = 1} \sum^C_{c=1}{log(\hat{p}_n^c, l_n^c)-2\times\frac{\sum{(\hat{p}_n^c \cap l_n^c)}}{\sum{\hat{p}_n^c} + \sum{l_n^c}} + 1},
\end{array}
\label{eq:seg}
\end{equation}
where $\hat{p}_n^c$ and $l_n^c$ denote the predicted and ground-truth segmentation masks for the $n$-th patch in the $c$-th category. In this work, we segment $c=7$ categories, including five parts of lung lobes and the background. It is worth noting that most of the cases in our dataset do not have ground-truth segmentation masks. For these cases, we simply avoid calculating the segmentation loss for them. 

Finally, the losses in Eqs.~\ref{eq:mil} and \ref{eq:seg} are trained simultaneously in a multi-task learning manner, and  the overall loss of the proposed method is written as
\begin{equation}
\begin{array}{rrclcl}
\displaystyle \mathcal{L} = \lambda \mathcal{L}_{MIL} +  \mathcal{L}_{Seg},
\end{array}
\label{eq:Ratio}
\end{equation} 
where $\lambda$ is the trade-off parameter used to balance the contributions of these two tasks. In this work, $\lambda$ is empirically set to $0.01$.

\begin{table*}[htbp]
\renewcommand{\arraystretch}{1.3}
\centering
\caption{Quantitative comparison for severity assessment tasks with the state-of-the-art methods.}
\setlength{\tabcolsep}{8pt}
\begin{tabular}{|c|c|c|c|c|c|}
\toprule[1pt]
\textbf{Method} & Accuracy & Precision & Recall & F1 Score & AUC \\
\toprule[1pt]
\textbf{ResNet50+Max~\cite{szegedy2017inception}} & 0.924 $\pm$ 0.497 & 0.856 $\pm$ 0.154 & 0.793 $\pm$ 0.106 & 0.816$\pm$ 0.120 & 0.803 $\pm$ 0.090 \\
\hline
\textbf{Gated Att. MIL~\cite{ATTMIL}} & 0.955 $\pm$ 0.015 & 0.879 $\pm$ 0.054 & 0.946 $\pm$ 0.019 & 0.906$\pm$ 0.037 & 0.973 $\pm$ 0.024 \\
\hline
\textbf{Tang~\etal~\cite{tang2020severity}}* & 0.875 & - & 0.933 & - & 0.910 \\
\hline
\textbf{Yang~\etal~\cite{yang2020severe}} & - & - & 0.750 & - & 0.892 \\
\hline
\textbf{Cls. Only (Ours)} & 0.969 $\pm$ 0.023 & 0.928 $\pm$ 0.073 & \textbf{0.958} $\pm$ \textbf{0.031} & 0.938 $\pm$ 0.045 & 0.980 $\pm$ 0.013 \\
\hline
\textbf{M$^2$UNet (Ours)} & \textbf{0.985} $\pm$ \textbf{0.005} & \textbf{0.975} $\pm$ \textbf{0.022} & 0.952 $\pm$ 0.011 & \textbf{0.963} $\pm$ \textbf{0.011} & \textbf{0.991} $\pm$ \textbf{0.010} \\
\toprule[1pt]
\end{tabular}\\
\label{Table:sota_class} 
\end{table*}

\begin{table*}[htbp]
\renewcommand{\arraystretch}{1.3}
\centering
\caption{Quantitative comparison for the performance of lung lobe segmentation with the state-of-the-art methods.}
\setlength{\tabcolsep}{8pt}
\begin{tabular}{|c|c|c|c|c|}
\toprule[1pt]
\textbf{Method} & \# (MB) & DSC & SEN & PPV \\
\toprule[1pt]
\textbf{U-Net} & $131.71$ & 0.776 $\pm$ 0.050 & 0.759 $\pm$ 0.037 & \textbf{0.834} $\pm$ \textbf{0.033}  \\
\hline
\textbf{U-Net++} & $34.97$ & 0.784 $\pm$ 0.035 & 0.773 $\pm$ 0.038 & 0.821 $\pm$ 0.018 \\
\hline
\textbf{Seg. Only (Ours)} & $14.37$ & 0.759 $\pm$ 0.055 & 0.756 $\pm$ 0.064 & 0.785 $\pm$ 0.045 \\
\hline
\textbf{M$^2$UNet (Ours)} & $15.32$ & \textbf{0.785} $\pm$ \textbf{0.058} & \textbf{0.783} $\pm$ \textbf{0.059} & 0.799 $\pm$ 0.051 \\
\toprule[1pt]
\end{tabular}\\
\label{Table:sota_seg}
\end{table*}

\subsection{Implementation}
The proposed method is implemented based on the open-source deep learning library \emph{Pytorch}. The training of the network is accelerated by four NVidia Tesla V100 GPUs (each with 32 GB memory). For feasible learning of the lung region images, we clamp the intensities of the image into $[-1200,0]$, which indicates that we use the width of $1200$ and the level of $-600$ for the pulmonary window. Then, the data is normalized to the value of $[0,255]$, as used by other deep learning methods. The dataset is highly imbalanced as the number of severe patients is much fewer than the non-severe patient. The ratio of the severe patient is less than $20\%$ in our dataset. Therefore, we augmented the data by directly duplicated the severe cases in the training set. This can be done because the proposed method uses a random cropping strategy to construct the inputs. This makes the duplicated cases not the same to each other for the training of the network. 
We also use the random cropping strategy in the testing stage, by assuming that the data distribution is already well learned in training. Other cropping strategies, \eg, center cropping, may not be suitable here, as the center of pulmonary is dominated by the trachea and other tissues.

In both training and testing stage, we randomly crop $200$ image patches from each input 3D CT image to construct the image-level bag (\ie, with the bag size of $n=200$).
And we use the output of the encoder to construct the embedding-level bag that contains $8\times8$ feature maps. 
We train M$^2$UNet using the learning rate of $0.01$ with a decay strategy of `Poly' (with the power of $0.75$). The network is optimized by a standard Stochastic Gradient Descent (SGD) algorithm with $100$ epochs. And the weights are decayed by a rate of $1\times10^{-4}$ with the momentum of $0.9$.

\section{Experiments}
In this section, we first introduce the materials, competing methods, and experimental setup. We then present the experimental results achieved by our method and several state-of-the-art methods. We finally investigate the influence of two major strategies used in our method. More results on parameter analysis can be found in the \emph{Supplementary Materials}.

\subsection{Materials}
The real COVID-19 dataset contains a total of $666$ 3D chest CT scans acquired from $242$ patients who are confirmed with COVID-19 (\ie, RT-PCR Test Positive). These CT images are collected from seven hospitals with a variety of CT scanners, including Philips (Ingenuity CT iDOSE4), GE (Bright speed S), Siemens (Somatom perspective), Hitachi (ECLOS), and Anke (ANATOM 16HD). The images are of large variation in terms of the image size of $512\times(512\sim666)\times(23\sim732)$, and the spatial resolution of $0.586\sim0.984\,mm$, $0.586\sim0.984\,mm$ and $0.399\sim10\,mm$. Obviously, diagnosis based on these images is a very challenging task.
The severity of the patient is confirmed by clinicians, following the guideline of 2019-nCoV (trail version 7) issued by the China National Health Commission. The severity of the patient is categorized into four types, \ie, mild, moderate, severe, and critical. In clinical practice, patients are often divided into two groups with different treatment regimens, \ie, severe and non-severe. 
The segmentation of 152 out of 666 images were delineated by an AI-based software and confirmed by experienced radiologists.
In this work, we employ this partitioning strategy. That is, mild and moderate are treated as non-severe, while severe and critical are regarded as severe. Therefore, the task of severe assessment is formulated into a binary classification problem.
Therefore, the dataset is partitioned into $51$ severe and $191$ non-severe patients.

\subsection{Competing Methods}

We first compare the proposed M$^2$UNet with four state-of-the-art methods in~\cite{tang2020severity,yang2020severe,szegedy2017inception,ATTMIL} for severity assessment of COVID-19 patients. The first two methods~\cite{tang2020severity,yang2020severe} are both based on hand-crafted features of CT images, while the last two~\cite{szegedy2017inception,ATTMIL} are deep learning-based methods that can learn imaging features automatically from data. Specifically, Tang~\etal~\cite{tang2020severity} first segment the lung, lung lobe and lesions in CT images. Then, the quantitative features of COVID-19 patients, e.g., the infection volume and the ratio of the whole lung, are calculated based on the segmentation results. The prediction is done by a random forest method. Yang~\etal~\cite{yang2020severe} propose to aggregate infection scores calculated on 20 lung regions for severity assessment.
Besides, the ResNet50+Max method~\cite{szegedy2017inception} is also compared for patch-wise classification. ResNet50+Max is a non-MIL method, which has a ResNet-50 network architecture and performs image-level classification through max-voting. In addition, we apply the Gated Att. MIL method proposed in~\cite{ATTMIL} on our dataset, which is a one-stage MIL method with an attention mechanism. For fair comparison, this method shares the same multi-instance pool as our M$^2$UNet.

We further compare our method with two state-of-the-art methods for lung lobe segmentation, including 1) UNet~\cite{ronneberger2015u}, and 2) UNet++~\cite{zhou2018unet++}. The parameter settings for these five competing methods are the same as those in their respective papers.

To evaluate the influence of the proposed multi-task learning and hierarchical MIL strategies used in M$^2$UNet, we further compare M$^2$UNet with its two variants: 1) M$^2$UNet with only the classification sub-network (denoted as Clas. Only), 2) M$^2$UNet with only the segmentation sub-network (denoted as Seg. Only). 

\begin{figure*}[!t]
  \centering
  \includegraphics[width=0.9\linewidth]{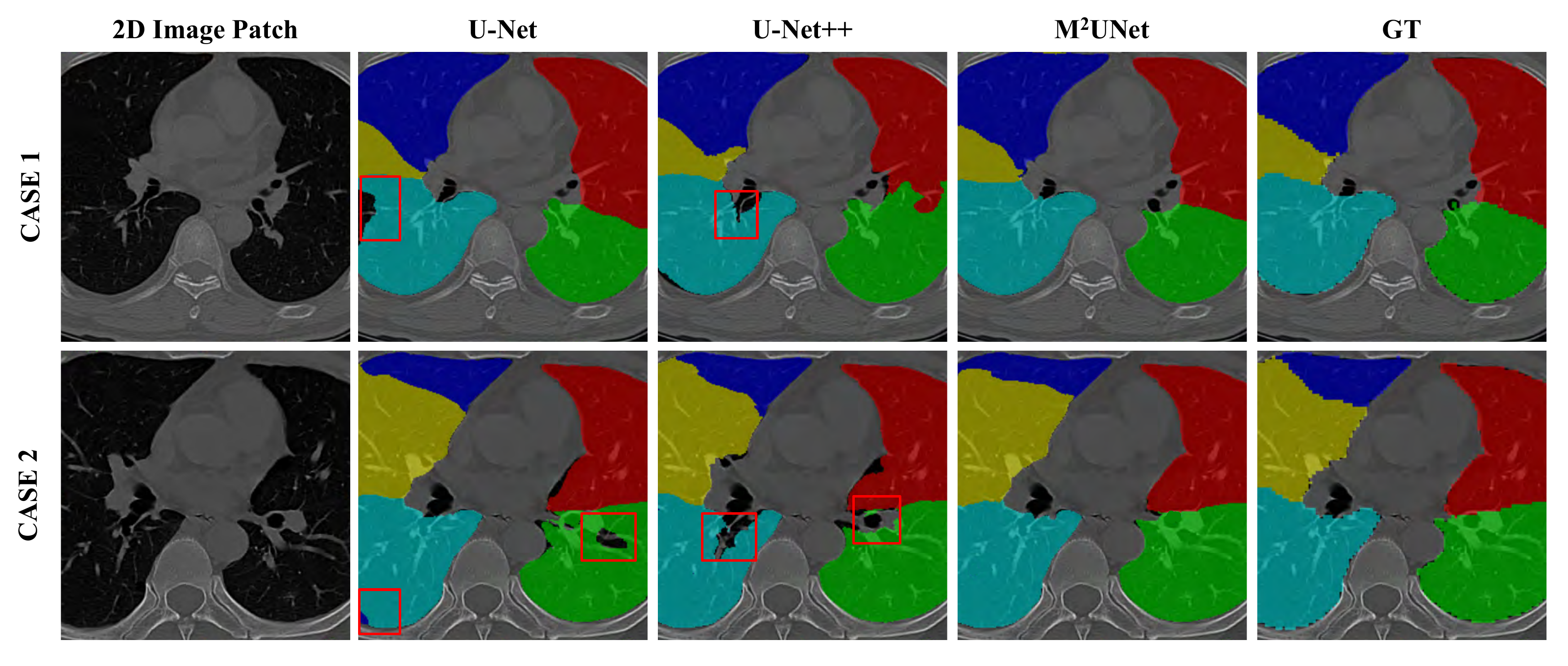}
    \caption{\label{fig:seg}The visualization of lung lobe segmentation results by three different methods on two typical cases. GT denotes the ground-truth masks. The under-segmentation regions are denoted by red boxes.}
\end{figure*}

\begin{table*}[!tbp]
\renewcommand{\arraystretch}{1.3}
\centering
\caption{Evaluation of the proposed hierarchical multi-instance learning strategy.}
\setlength{\tabcolsep}{8pt}
\begin{tabular}{|c|c|c|c|c|c|}
\toprule[1pt]
\textbf{Method} & Accuracy & Precision & Recall & F1 Score & AUC \\
\toprule[1pt]
\textbf{ResNet50+Max~\cite{szegedy2017inception}} & 0.924 $\pm$ 0.497 & 0.856 $\pm$ 0.154 & 0.793 $\pm$ 0.106 & 0.816$\pm$ 0.120 & 0.803 $\pm$ 0.090 \\
\hline
\textbf{Gated Att. MIL~\cite{ATTMIL}} & 0.955 $\pm$ 0.015 & 0.879 $\pm$ 0.054 & 0.946 $\pm$ 0.019 & 0.906$\pm$ 0.037 & 0.973 $\pm$ 0.024 \\
\hline
\textbf{Cls. Only (Ours)} & \textbf{0.969} $\pm$ \textbf{0.023} & \textbf{0.928} $\pm$ \textbf{0.073} & \textbf{0.958} $\pm$ \textbf{0.031} & \textbf{0.938}$\pm$ \textbf{0.045} & \textbf{0.980} $\pm$ \textbf{0.013} \\
\toprule[1pt]
\end{tabular}\\
\label{Table:mil} 
\end{table*}
\begin{table*}[!tbp]
\renewcommand{\arraystretch}{1.3}
\centering
\caption{Evaluation of the proposed multi-task learning strategy for severity assessment.}
\setlength{\tabcolsep}{8pt}
\begin{tabular}{|c|c|c|c|c|c|}
\toprule[1pt]
\textbf{Method} & Accuracy & Precision & Recall & F1 Score & AUC \\
\toprule[1pt]
\textbf{Cls. Only} & 0.969 $\pm$ 0.023 & 0.928 $\pm$ 0.073 & \textbf{0.958} $\pm$ \textbf{0.031} & 0.938 $\pm$ 0.045 & 0.980 $\pm$ 0.013 \\
\hline
\textbf{M$^2$UNet} & \textbf{0.985} $\pm$ \textbf{0.005} & \textbf{0.975} $\pm$ \textbf{0.022} & 0.952 $\pm$ 0.011 & \textbf{0.963} $\pm$ \textbf{0.011} & \textbf{0.991} $\pm$ \textbf{0.010} \\
\toprule[1pt]
\end{tabular}\\
\label{Table:cls}
\end{table*}

\subsection{Experimental Setup}
A five-fold cross-validation (CV) strategy is used in the experiments for performance evaluation. Specifically, the whole dataset is first randomly partitioned into five subsets (with approximately equal sample size of subjects). We treat one subset as the testing set ($20\%$), while the remaining four subsets are combined to construct the training set ($70\%$) and validation set ($10\%$). The validation set here is used for selecting the hyper-parameters. This process is iterated until each subsets serve as a testing set once. The final results are reported on the test set.

Two tasks are included in the proposed method, \ie, classification of severity assessment, and 2) segmentation of the lung lobe. For performance evaluation, two sets of metrics are used in these two tasks, with the details given below.

\subsubsection{Metrics for Classification}
We use five commonly used metrics to evaluate the classification performance achieved by different methods in the severity assessment task, \ie, Accuracy, Precision, Recall, F1 Score, and the area under the receiver operating characteristic curve (AUC).
\begin{equation}
\footnotesize
\begin{array}{rrclcl}
\displaystyle Accuracy= \frac {TP + TN}{TP + TN + FP + FN},
\end{array}
\label{eq:Accuracy} 
\end{equation}

\begin{equation}
\footnotesize
\begin{array}{rrclcl}
\displaystyle Precision=\frac{TP}{TP + FP},
\end{array}
\begin{array}{rrclcl}
\displaystyle Recall=\frac{TP}{TP + FN},
\end{array}
\label{eq:PPV} 
\end{equation}

\begin{equation}
\footnotesize
\begin{array}{rrclcl}
\displaystyle F1~Score = \frac {2 \times Precision \times Recall}{Precision + Recall}.
\end{array}
\label{eq:F1} 
\end{equation}
where TP, TN, FP and FN denote true positive, true negative, false positive and false negative, respectively.

\subsubsection{Metrics for Segmentation}
We use three metrics, \ie, Dice Similarity Coefficient (DSC), Positive Predict Value (PPV) and Sensitivity (SEN), to evaluate the segmentation performance of different methods, with the definitions given below.
\begin{equation}
\footnotesize
\begin{array}{rrclcl}
\displaystyle DSC=\frac{2\|V_{gt}\cap V_{seg}\|}{\|V_{gt}\|+\|V_{seg}\|};
\end{array}
\label{eq:DSC} 
\end{equation}
\begin{equation}
\footnotesize
\begin{array}{rrclcl}
\displaystyle PPV=\frac{\|V_{gt} \cap V_{seg}\|}{\|V_{seg}\|};~~~~~~~~~~SEN=\frac{\|V_{gt} \cap V_{seg}\|}{\|V_{gt}\|}.
\end{array}
\end{equation}

where $V_{gt}$ and $V_{seg}$ represent the ground-truth and predicted segmentation maps for each scan, respectively.

\begin{figure}[!t]
  \centering
  \includegraphics[width=\linewidth]{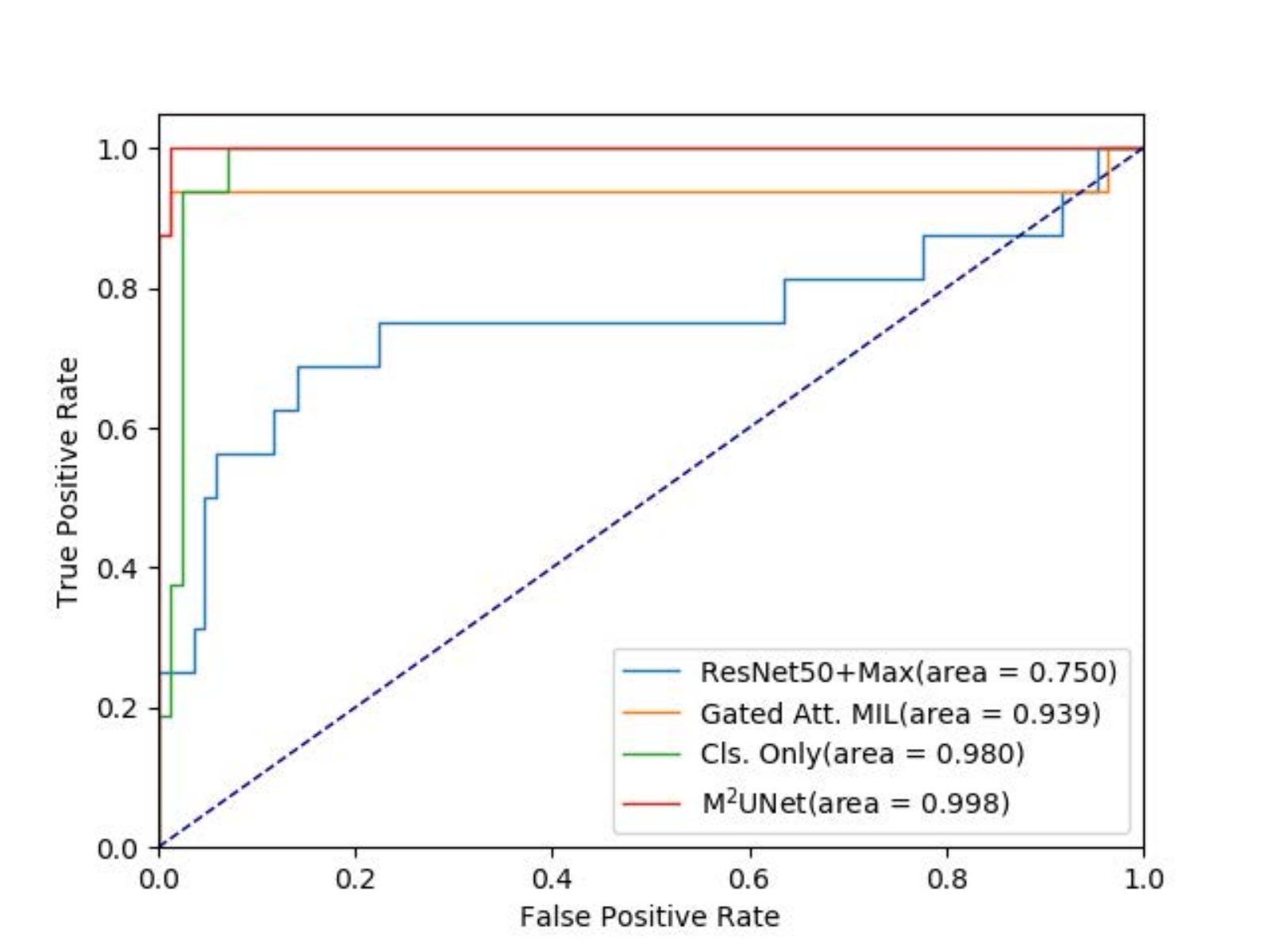}
    \caption{\label{fig:roc}The receiver operating characteristic (ROC) curves achieved by four different methods in the task of severity assessment.}
\end{figure}

\subsection{Comparison with State-of-the-art Methods}
\subsubsection{Results of Severity Assessment}
We first report the performance of six different methods in the task of severity assessment for COVID-19 patients, with the results shown in Table~\ref{Table:sota_class}. Note that the results from the competing methods are directly obtained from the respective papers. As can be seen, four deep learning-based methods usually outperform two hand-crafted feature-based methods in most cases. For some specific metrics, the method in \cite{tang2020severity} achieves the Recall of $0.933$, which is significantly better than the non-MIL method ResNet50+Max. The conventional MIL-based method in~\cite{ATTMIL} gets a performance improvement in terms of accuracy by $8\%$. 
Three MIL methods (\ie,~\cite{ATTMIL}, Cls. Only, and M$^2$UNet) yield satisfying performance, and the proposed M$^2$UNet achieves the best results (\eg., the accuracy of $98.5\%$ and F1 Score of $99.1\%$). 
However, the proposed method with multiple instances in multi-instance learning achieves the accuracy of $98.5\%$ and F1 Score of $99.1\%$.
The receiver operating characteristic (ROC) curves of six competing methods are illustrated in Fig.~\ref{fig:roc}. Note that this ROC curve is plotted based on the results on one fold testing data, which is slightly different from the average performance on five-folds in Table~\ref{Table:sota_class}. Table~\ref{Table:sota_class} and Fig.~\ref{fig:roc} clearly suggest that our M$^2$UNet generates the overall best performance in the task of severity assessment of COVID-19 patients based on chest CT images.

\subsubsection{Results of Lung Lobe Segmentation}
We then report the results of lung lobe segmentation achieved by four different methods in Table~\ref{Table:sota_seg}. Comparing Seg. Only and the conventional U-Net, the former dramatically reduces the parameter from $131.71 \textrm{MB}$ to $14.37 \textrm{MB}$. As a consequence, the performance in terms of DSC and PPV is also decreased by $1.7\%$ and $4.9\%$, respectively. By using multi-task learning, M$^2$Net improves the performance, from $0.759$ to $0.785$ in terms of DSC, which also outperform the performance of conventional U-Net, with a decreasing of parameters, from $131.71$ to $15.32$. The proposed M$^2$UNet also achieves a slightly higher performance compared with U-Net++.

The visualization of segmentation results achieved by three different methods on two subjects is shown in Fig.~\ref{fig:seg}. From this figure, we can see that M$^2$UNet generates the overall best segmentation masks, while U-Net and U-Net++ usually yield under-segmentation results on these cases. These results further show the advantage of our M$^2$UNet.

\begin{table*}[!htbp]
\renewcommand{\arraystretch}{1.3}
\centering
\caption{\label{Table:bagsize} Performance comparison for severity prediction with respect to different bag size.}
\setlength{\tabcolsep}{8pt}
\begin{tabular}{|c|c|c|c|c|}
\toprule[1pt]
 Bag Size & Accurate & Precision & Recall & F1 Score \\
\toprule[1pt]
 $50$ & 0.933 $\pm$ 0.045 & 0.845 $\pm$ 0.072 & 0.936 $\pm$ 0.043 & 0.878 $\pm$ 0.060 \\
 \hline
 $80$ & 0.940 $\pm$ 0.030 & 0.844 $\pm$ 0.069 & 0.948 $\pm$ 0.037 & 0.882 $\pm$ 0.060 \\
 \hline
 $100$ & 0.965 $\pm$ 0.032 & 0.914 $\pm$ 0.061 & 0.959 $\pm$ 0.042 & 0.931 $\pm$ 0.048 \\
 \hline
 $150$ & \textbf{0.971} $\pm$ \textbf{0.019} & \textbf{0.929} $\pm$ \textbf{0.044} & 0.946 $\pm$ 0.036 & 0.935 $\pm$ 0.029 \\
 \hline
 $200$ & 0.969 $\pm$ 0.023 & 0.928 $\pm$ 0.073 & \textbf{0.958} $\pm$ \textbf{0.031} & \textbf{0.938} $\pm$ \textbf{0.045} \\
\toprule[1pt]
\end{tabular}\\
\end{table*}

\subsection{Ablation Study}
We further evaluate the influence of two major strategies used in our M$^2$UNet, \ie, 1) the hierarchical MIL strategy for classification, and 2) the multi-task learning strategy for joint severity assessment and lung lobe segmentation.

\subsubsection{Influence of Hierarchical MIL Strategy}
To evaluate the effectiveness of the hierarchical MIL strategy, we compare the variant of the proposed M$^2$UNet (\ie, Cls. Only without the segmentation subnetwork) with a non-MIL method (\ie, ResNet50+Max) and a one-stage MIL method (\ie, Gated Att. MIL~\cite{ATTMIL}). 
The classification results of these three methods in the task of severity assessment are reported in Table~\ref{Table:mil}. 
As shown in Table~\ref{Table:mil}, two MIL methods (\ie, Gated Att. MIL and Cls. Only) can generate more accurate decisions under the weakly supervised setting, compared with the non-MIL method ResNet50+Max. Besides, our hierarchical MIL strategy can further boost the classification performance compared to the conventional one-stage MIL strategy. For instance, our Cls. Only method achieves an F1 Score of $0.938$, which is higher than that (\ie, $0.906$) yielded by Gated Att. MIL with a one-stage MIL strategy. These results suggest the effectiveness of the proposed hierarchical MIL strategy.

\subsubsection{Influence of Multi-task Learning Strategy}
Our M$^2$UNet can jointly learn the segmentation task and the classification task in a multi-task learning manner. Here, we also investigate the influence of such a multi-task learning paradigm, by comparing M$^2$UNet with its two single-task variants, \ie, `Cls. Only' for classification and `Seg. Only' for segmentation.
The performance comparison in two tasks for severity assessment and lung lobe segmentation are reported in Table~\ref{Table:cls} and Table~\ref{Table:seg}, respectively. Table~\ref{Table:cls} suggests that, compared with Cls. Only, the multi-task learning paradigm used in M$^2$UNet helps to improve the classification accuracy by $1.6\%$, while increasing the precision score by over $5\%$ and the F1 Score by $2.5\%$. Notably, the F1 and precision of the Cls. Only method are already higher than $90\%$, which are hard to be improved. This is more valuable in this classification scheme, as the F1 score is more representative in evaluating such an imbalanced classification task.

As can be observed from Table~\ref{Table:seg}, although M$^2$UNet is not specifically designed for lung lobe segmentation, it still improves the overall segmentation performance in terms of three metrics, compared with its single-task variant (\ie, Seg. Only). This implies that the proposed multi-task learning strategy is useful in boosting the learning performance of both tasks of severity assessment and lung lobe segmentation.

\begin{table}[!tbp]
\renewcommand{\arraystretch}{1.3}
\centering
\caption{Evaluation of the proposed multi-task learning strategy for lung lobe segmentation.}
\setlength{\tabcolsep}{5pt}
\begin{tabular}{|c|c|c|c|c|}
\toprule[1pt]
\textbf{Method}& DSC & SEN & PPV \\
\toprule[1pt]
\textbf{Seg. Only} & 0.759 $\pm$ 0.055 & 0.756 $\pm$ 0.064 & 0.785 $\pm$ 0.045 \\
\hline
\textbf{M$^2$UNet} & \textbf{0.785} $\pm$ \textbf{0.058} & \textbf{0.783} $\pm$ \textbf{0.059} & \textbf{0.799} $\pm$ \textbf{0.051} \\
\toprule[1pt]
\end{tabular}\\
\label{Table:seg}
\end{table}

\subsection{Influence of Bag Size}

We further investigate the performance of our method using different size of bags, and the results are shown in Table~\ref{Table:bagsize}. Specifically, we vary the bag size within $\{50,80,100,150,200\}$. As shown in the table, the performance of M$^2$UNet for classification changes along with the bag size. The proposed method achieves the best performance with the bag size of $200$. Another observation is that, the table suggests the performance of the proposed AI-based severity assessment model is not sensitive with the bag size when larger than $100$, indicating that at least $100$ patches are required for the proposed method to achieve an acceptable result.

\section{Conclusion and Future Work}
In this paper, we propose a synergistic learning framework for automated severity assessment and lung segmentation of COVID-19 in 3D chest CT images. In this framework, we first represent each input image by a bag to deal with the challenging problem that the severity is related to local infected regions in the CT image. We further develop a multi-task multi-instance deep network (called M$^2$UNet) to assess the severity of COVID-19 patients and segment the lung lobe simultaneously, where the context information provided by segmentation can be employed to boost the performance of severity assessment. A hierarchical multi-instance learning strategy is also proposed in M$^2$UNet for severity assessment. Experimental results on a real COVID-19 CT image dataset demonstrate that our method achieves promising results in severity assessment of COVID-19 patients, compared with several state-of-the-art methods.

In the current work, the severity assessment of COVID-19 only relies on one time-point data, without considering longitudinal imaging biomarkers. It is interesting to perform a longitudinal study to investigate the progression of the disease, which is one of our future work. 
Since annotations for lung lobe in 3D CT images are usually tedious and error-prone, only a small number of subjects in our dataset have ground-truth segmentation. Therefore, it is highly desired to develop automated or even semi-automated image annotation methods, which will also be studied in the future.




\ifCLASSOPTIONcaptionsoff
  \newpage
\fi

\bibliographystyle{IEEEtran}
\bibliography{IEEEabrv,covtmi}

\end{document}


\title{Synergistic Learning of Lung Lobe Segmentation and Hierarchical Multi-Instance Classification for Automated Severity Assessment of COVID-19 in CT Images\\
-- \emph{Supplementary Materials}} 

\maketitle

Besides the experimental results in the main text, we further analyze the influence of parameters on the performance of the proposed M$^2$UNet, with the details given below.


\section{Influence of Weigh Parameter}
\begin{figure}[htbp]
  \centering
  \includegraphics[width=\linewidth]{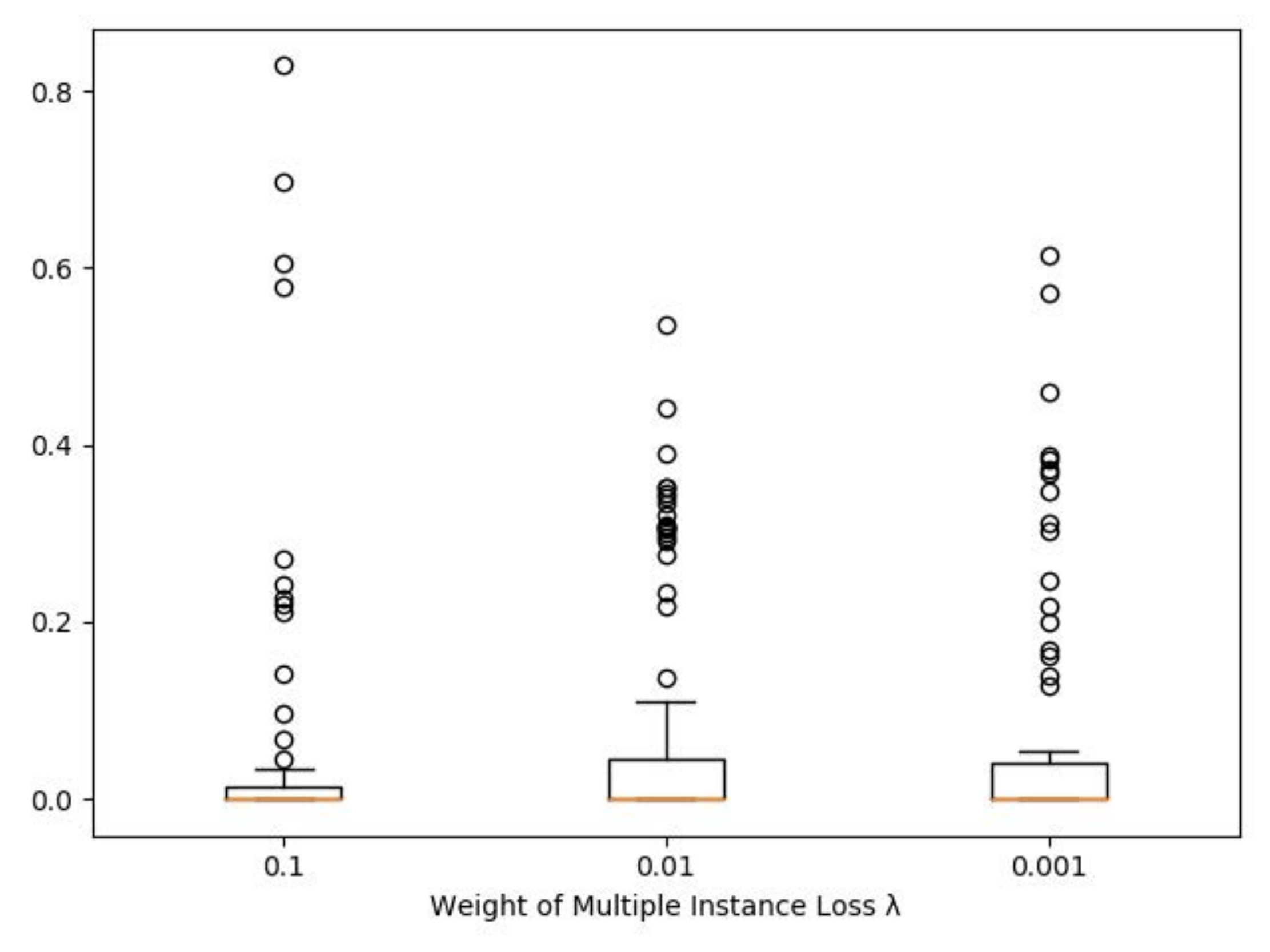}
    \caption{\label{fig:box}Boxplot for the margin of prediction and label with respect to different weight $\lambda$ for multiple instance loss. The margin larger than $0.5$ indicates the wrong prediction.}
\end{figure}

We evaluate the influence of the hyper-parameter $\lambda$. The boxplot for the margin of prediction and ground-truth labels is shown in Fig.~\ref{fig:box}, where the margin is calculated by $|p-l|$, $p$ denotes the prediction score and $l$ denotes the label. The margin below $0.5$ indicates that the prediction is correct. As shown in this figure, the task weight $\lambda$ affects the classification performance of the model, and the proposed M$^2$UNet performs best with $\lambda=0.01$. 

\section{Influence of Learning Rate}
We further investigate the influence of the learning rate on the performance of M$^2$UNet, with the results given in Tables~\ref{Table:lr_seg}-\ref{Table:lr_cls}. As suggested by the table, the performance of M$^2$UNet for both classification and segmentation tasks are affected by different learning rates. The proposed method achieves the best performance for both classification and segmentation tasks with the learning rate of $0.01$. Another observation is that the performance of the network is unstable with a large learning rate (i.e., 0.1), and the gap between precision and recall is large, with $0.833$ in terms of precision and $0.970$ in terms of recall. The performance of the network is stable with a small learning rate (i.e., 0.001), with precision of $0.881$ and recall of $0.922$. However, it cannot achieve the best performance, compared with the network trained by the learning rate of $0.01$.

\begin{table}[!htbp]
\renewcommand{\arraystretch}{1.3}
\centering
\caption{\label{Table:lr_seg} Performance comparison for segmentation with respect to different learning rate.}
\setlength{\tabcolsep}{8pt}
\begin{tabular}{|c|c|c|c|}
\toprule[1pt]
 Learning rate & DSC & SEN & PPV \\
\toprule[1pt]
$0.1$ & 0.783  & 0.772 & 0.789 \\
\hline
$0.01$ & 0.785  & 0.783 & 0.799 \\
\hline
$0.001$ & 0.734  & 0.705 & 0.754 \\
\toprule[1pt]
\end{tabular}\\
\end{table}

\begin{table}[!htbp]
\renewcommand{\arraystretch}{1.3}
\centering
\caption{\label{Table:lr_cls} Performance comparison for severity prediction with respect to different learning rate.}
\setlength{\tabcolsep}{8pt}
\begin{tabular}{|c|c|c|c|c|}
\toprule[1pt]
 Learning rate & Accurate & Precision & Recall & F1 Score \\
\toprule[1pt]
 $0.1$ & 0.946 & 0.833 & 0.970 & 0.885 \\
 \hline
 $0.01$ & 0.985 & 0.975 & 0.952 & 0.963 \\
 \hline
 $0.001$ & 0.960 & 0.881 & 0.922 & 0.900 \\
\toprule[1pt]
\end{tabular}\\
\end{table}
